\documentclass[sigconf]{acmart}
\AtBeginDocument{%
  }

\copyrightyear{2025}
\acmYear{2025}
\setcopyright{rightsretained}
\acmConference[KDD '25] {Proceedings of the 31st ACM SIGKDD Conference on Knowledge Discovery and Data Mining V.1}{August 3--7, 2025}{Toronto, ON, Canada.}
\acmBooktitle{Proceedings of the 31st ACM SIGKDD Conference on Knowledge Discovery and Data Mining V.1 (KDD '25), August 3--7, 2025, Toronto, ON, Canada}
\acmISBN{979-8-4007-1245-6/25/08}
\acmDOI{10.1145/3690624.3709429}

\usepackage{amsmath,amsfonts,bm,mathtools}

\def\eqref#1{equation~\ref{#1}}

\def\1{\bm{1}}

\DeclareMathAlphabet{\mathsfit}{\encodingdefault}{\sfdefault}{m}{sl}
\SetMathAlphabet{\mathsfit}{bold}{\encodingdefault}{\sfdefault}{bx}{n}

\DeclareMathOperator*{\argmax}{arg\,max}

\def\L{\mathcal{L}}

\usepackage{amsmath}
\usepackage{algorithm,algorithmic}
\renewcommand{\algorithmiccomment}[1]{\bgroup\hfill//~#1\egroup}

\usepackage{subcaption}
\usepackage{multirow}
\usepackage{array}

\begin{document}


\title{Beyond Item Dissimilarities: Diversifying by Intent in Recommender Systems} 

\author{Yuyan Wang}
\email{yuyanw@stanford.edu}
\orcid{0000-0001-5585-3789}
\affiliation{%
  \institution{Stanford University}
  \streetaddress{655 Knight Way}
  \city{Stanford}
  \state{CA}
  \country{USA}
  \postcode{94305}
}

\author{Cheenar Banerjee}
\email{cheenar@google.com}
\affiliation{%
  \institution{Google, Inc.}
  \streetaddress{1600 Amphitheatre Parkway}
  \city{Mountain View}
  \state{CA}
  \country{USA}
  \postcode{94043}
}

\author{Samer Chucri}
\email{samerc@google.com}
\affiliation{%
  \institution{Google, Inc.}
  \streetaddress{1600 Amphitheatre Parkway}
  \city{Mountain View}
  \state{CA}
  \country{USA}
  \postcode{94043}
}

\author{Fabio Soldo}
\email{fsoldo@google.com}
\affiliation{%
  \institution{Google, Inc.}
  \streetaddress{1600 Amphitheatre Parkway}
  \city{Mountain View}
  \state{CA}
  \country{USA}
  \postcode{94043}
}

\author{Sriraj Badam}
\email{srirajdutt@google.com}
\affiliation{%
  \institution{Google, Inc.}
  \streetaddress{1600 Amphitheatre Parkway}
  \city{Mountain View}
  \state{CA}
  \country{USA}
  \postcode{94043}
}

\author{Ed H. Chi}
\email{edchi@google.com}
\affiliation{%
  \institution{Google DeepMind}
  \streetaddress{1600 Amphitheatre Parkway}
  \city{Mountain View}
  \state{CA}
  \country{USA}
  \postcode{94043}
}

\author{Minmin Chen}
\email{minminc@google.com}
\affiliation{%
  \institution{Google DeepMind}
  \streetaddress{1600 Amphitheatre Parkway}
  \city{Mountain View}
  \state{CA}
  \country{USA}
  \postcode{94043}
}


\renewcommand{\shortauthors}{Yuyan Wang et al.}

\begin{abstract}

It has become increasingly clear that recommender systems that overly focus on short-term engagement prevents users from exploring diverse interests, ultimately hurting long-term user experience. To tackle this challenge, numerous diversification algorithms have been proposed as the final stage of recommender systems. These algorithms typically rely on measures of item similarity, aiming to maximize the dissimilarity across items in the final set of recommendations. However, in this work, we demonstrate the benefits of going beyond item-level similarities by utilizing higher-level user understanding—specifically, user intents that persist across multiple interactions or recommendation sessions—in diversification. Our approach is motivated by the observation that user behaviors on online platforms are largely driven by their underlying intents. Therefore, recommendations should ensure that diverse user intents are accurately represented. While intent has primarily been studied in the context of search, it is less clear how to incorporate real-time dynamic intent predictions into recommender systems.

To address this gap, we develop a probabilistic intent-based whole-page diversification framework for the final stage of a recommender system. Starting with a prior belief of user intents, the proposed framework sequentially selects items for each position based on these beliefs and subsequently updates posterior beliefs about the intents. This approach ensures that different user intents are represented on a page, towards optimizing long-term user experience.

We experiment with the intent diversification framework on YouTube, the world's largest video recommendation platform, serving billions of users daily. Live experiments on a diverse set of intents show that the proposed framework increases Daily Active Users (DAU) and overall user enjoyment, validating its effectiveness in facilitating long-term planning. Specifically, it enables users to consistently discover and engage with diverse content that aligns with their underlying intents over time, leading to an improved long-term user experience.

\end{abstract}




\maketitle


\section{Introduction}
\label{sec:intro}

Recommender systems, one of the biggest successes of big data and machine learning in industry applications, have become an integral part of the user experience on online platforms. Traditional recommender systems that focus on optimizing users’ immediate responses such as clicks, likes and dwell time in the current session have gained tremendous success over the years \citep{adomavicius2005toward, covington2016deep}, setting the foundation for personalization \citep{zhang2019deep}. 

However, it has become increasingly clear that focusing on optimizing short-term engagement can inadvertently lead to undesirable outcomes and hurt long-term user experience. In particular, this can lead to pigeon-holing and echo-chamber effects \citep{chaney2018algorithmic, nguyen2014exploring} that prevent users from exploring new interests, resulting in a bad long-term user experience. To tackle this challenge, numerous exploration and diversification algorithms have been proposed to improve the novelty and diversity of the recommendation results \citep{carbonell1998use, chen2021values, wilhelm2018practical}. These algorithms rely item-level notions such as novelty \citep{herlocker2004evaluating, chen2021values} or pairwise similarities \citep{carbonell1998use, ziegler2005improving}, aiming to maximize the dissimilarity across items in the final set of recommendations. 

In this work, we demonstrate the benefits of going beyond item-level similarities by utilizing a higher-order user understanding—specifically, user intents that persist across multiple interactions or recommendation sessions—in diversifying recommendation results. This is motivated by the observation that user behaviors on the online platforms are largely driven by their underlying \emph{intents} \citep{fishbein1977belief}. These intents can vary not only across different users but also within the same user over time. Therefore, an ideal diversification algorithm should account for the different intents that a user may have. In particular, it should diversify not only across topics but also across the different intents of the user, in order to provide coherent recommendation results and improve user experience in the long term.

Incorporating user intents in diversification has been primarily investigated within the context of search \citep{hu2011characterizing, cheng2010actively}, but remains largely under-explored for content recommendation platforms. To bridge this gap, in this work we develop an intent diversification framework that is applied to the final stage of a recommender system, by adapting a greedy algorithm originally proposed for search results diversification \citep{agrawal2009diversifying}. The diversification framework starts with a prior belief of user intents, represented by a probability distribution over different intents of the user, predicted by the intent prediction module. It sequentially selects items from the top position to the bottom, maximizing the probability of satisfying the user intent distribution at each step. After selecting each item at each position, it adjusts the posterior beliefs about user intents by assuming that the user is \emph{not} interested in the previously placed items, thereby ensuring that a diverse set of intents are accurately represented in the final recommendations. 

We experiment with our proposed intent diversification framework on YouTube, the world’s largest video recommendation platform, serving billions of users everyday. We instantiate the framework with a diverse set of user intents. Live A/B experiments on YouTube demonstrate that our intent diversification framework enables users to consistently discover and engage with diverse content that aligns with their underlying intents over time. Our framework leads to significant improvements in business metrics, including user retention (i.e., Daily Active Users, DAU) and overall user enjoyment, validating its effectiveness in optimizing long-term user experience.



\section{Related Work}
\label{sec:related}

\subsection{Diversification in Recommender Systems}
\label{sec:related_diversification}

Industrial recommender systems started as two-stage framework \citep{covington2016deep}: Candidate generation and ranking. The candidate generation stage provides broad personalization and narrows down millions of candidates to a few hundred. The ranking stage further selects a few dozen of items from the nominated candidates. Both stages rely on pointwise scores to select items that are of top relevance to the user. In other words, similarity among the selected items are ignored. However, seeing a page of very similar items however can be a suboptimal recommendation experience for the users as it may lead to echo-chamber and pigeon-holing effects, preventing the users from exploring diverse interests \citep{chaney2018algorithmic, ge2020understanding}. Recognizing this challenge, state-of-art recommender systems in recent years adopt a three-stage framework by adding a diversification or whole-page optimization stage as the final stage of the recommender system \citep{wilhelm2018practical, gao2023survey}, to ensure that a diverse set of contents is represented as the final recommendations \citep{ziegler2005improving}. This stage is a combinatorial optimization problem in nature as it needs to decide the optimal ordering of a set of candidates selected from the ranking stage. Several methods have been proposed over the years to reduce the complexity of the problem and achieve scalable diversification in large-scale systems, dating back to Carbonell’s work \citep{carbonell1998use} which proposed to greedily add items to the recommendation list with maximal marginal relevance (MMR). Recent works on diversification and whole-page optimization include leveraging determinantal point processes (DPP) \citep{wilhelm2018practical}, listwise modeling with recurrent or attentive sequence modeling \citep{zhuang2018globally, ai2018learning, pei2019personalized, huang2020personalized}, graph theories\citep{adomavicius2011maximizing}, and reinforcement learning \citep{zou2019reinforcement}. 

To our knowledge, most diversification algorithms to date still exclusively rely on \emph{item-level} similarity and dissimilarity measures, lacking a higher-level understanding of user intents. This indicates a missed opportunity for providing a coherent recommendation experience based on user intents that supports long-term user engagement and satisfaction. Our proposed intent diversification framework addresses this gap by integrating user intents into the final stage of a recommender system, ensuring that different user intents are accurately represented in the final recommendations towards a coherent and improved long-term user experience.

\subsection{User Intent Modeling and Application}
\label{sec:related_intent}

Intents have been widely studied across various fields including shopping \citep{lee2006shopping, morrison1979purchase}, advertising \citep{teixeira2014and}, online platforms \citep{ding2015learning}, employment \citep{guimaraes1992determinants} and public health \citep{serra2023incentives}. Psychological theories propose that people's intents are dynamic and subject to influence by various factors \citep{fishbein1977belief, ariely1998predictably}, making \emph{real-time} modeling necessary to accurately capture and respond to these dynamic intents.


User intent modeling has been largely explored within the search context \citep{nguyen2004capturing}. Broder's seminal research from over two decades ago ~\cite{broder2002taxonomy} proposed categorizing intents for a web search query into three predefined categories: informational, navigational, and transactional. These categories are widely utilized today for query understanding and query expansion \citep{azad2019query}. Numerous methods have been proposed to automatically extract or predict user goals and intents from a user's web query \citep{rose2004understanding, zhang2019generic, ashkan2009classifying, kong2015predicting, cheng2010actively}, which are then used to improve the quality of search engines' results, including the diversity of the search engine results page (SERP) \citep{chapelle2011intent, agrawal2009diversifying, santos2011intent}. 

While search and recommender systems adopt many similar methodologies, and user intent has been extensively researched in the search context, there appears to be a significant gap regarding user intent within recommender systems. Closer to our work is \cite{li2023variety} which incorporates users’ variety-seeking behavior as a feature in recommender systems, and \cite{ding2015learning} which uses HMMs to learn implicit user intents and decides dynamically the optimal web page adaption. However, \cite{li2023variety} assumes a static variety-seeking behavior for every user, therefore unable to capture the dynamic and evolving nature of the true underlying intent as our framework does. \cite{ding2015learning} is only able to optimize among a few actions and therefore not applicable to an industrial recommender system setting with billions of candidates. Our work aims at bridging this gap by proposing a scalable intent diversification framework that adapts to dynamic user intents.


\section{Method}
\label{sec:method}


\subsection{Notations and Definitions}
\label{sec:method_components}

We begin by introducing the components of the intent diversification framework, which is applied to the set of recommendation items selected by the nomination and ranking stages. We use $i$ to index users and $j$ to index recommendation items. Let there be $M$ items selected for the page, with scores denoted as $s_{i1}, ..., s_{iM}$, representing the outputs from the multi-stage recommender system. 

To introduce users' dynamic intents into diversification, we denote the space of intents as $\mathcal{V}$, with $v$ representing a specific intent within this space. Let $\Pr(v | i)$ denote the prior belief of user $i$'s propensity to have intent $v$ for the current page, which is the output of a classification model trained on past user behaviors as described below in Section \ref{sec:user_intent_modeling}. Let $Q (j | i)$ represent the value function of item $j$ for user $i$, which is readily available in existing intent-agnostic recommendation systems. We adopt a probabilistic definition of $Q (j | i)$, where it denotes the probability that user $i$ enjoys item $j$. This probabilistic definition is chosen so that its complement, $1 - Q (j | i)$, represents the probability that user $i$ does \emph{not} enjoy item $j$. As we will see shortly, this formulation facilitates the development of a \emph{probabilistic} framework for diversification, and is also widely adopted approach within search contexts \citep{agrawal2009diversifying}.

To diversify based on user intents, we also need an intent-conditioned value function $Q (j | i, v)$, which measures the probability that user $i$ enjoys item $j$, \emph{given} that she has intent $v$. By Bayes' theorem, the intent-conditioned value function can be written as 
\begin{equation}
\label{eq:intent_cond_val}
Q( j | i, v) = \frac{Q( j | i) \Pr(v | i, j) }{\Pr(v | i)}.
\end{equation}

The second term in the numerator, $\Pr(v | i, j)$, represents the probability that user $i$ has intent $v$, \emph{given} the consumption of item $j$. Intuitively, this value depends on whether item $j$ aligns with the user's intent $v$, denoted as $v \in \mathcal{V}_j$. By definition of intent, the user would not consume an item that is not aligned with her intents. Consequently, if item $j$ does not align with intent $v$, $\Pr(v | i, j) = 0$. 
On the other hand, if item $j$ aligns with intent $v$, then the item-level intent prediction $\Pr(v | i, j)$ equals the page-level intent prediction $\Pr(v | i)$ by law of total probability, assuming a user consumes at most one item in a page\footnote{$\Pr(v | i) = \sum_{j'} \Pr(v | i, j') \Pr(j' | i)$ and only one of the terms in the summation is nonzero: $\Pr(j | i)=1$ when user $i$ consumes item $j$.}. 
Therefore, Eq.(\ref{eq:intent_cond_val}) simplifies to 
\begin{equation}
\label{eq:intent_cond_result}
Q( j | i, v) = 
\begin{cases}
    Q( j | i),& \text{if } v \in \mathcal{V}_j,\\
    0,              & \text{otherwise}.
\end{cases}
\end{equation}

\subsection{User Intent Modeling}
\label{sec:user_intent_modeling}

An important input to our proposed intent diversification algorithm is the real-time intent probabilities $\Pr(v | i)$. We propose a supervised ML model $f^v(\cdot; \theta)$, parameterized by $\theta$, to generate a personalized, contextualized, and \emph{real-time} probability that user $i$ has intent $v$ in the current recommendation page. 
The input features, denoted by $x$, include user behavior signals such as past consumption patterns, session-level features such as session length and average completion ratio in the current session, and contextual features such as time of the day and day of the week.\footnote{For business compliance reasons, we omit the full feature list here.}

Identifying a user’s intent without directly asking her is a significant challenge. To address this, we define the intent label by leveraging behavior signals on the page. Specifically, we let the label $y^v = 1$ if the consumer consumes an item $j$ that aligns intent $v$ in the current recommendation page, i.e. $v \in \mathcal{V}_j$, and 0 otherwise. For example, we can define the user on a particular recommendation page to have an exploration intent if she consumes an exploration item on the page, where an exploration item is defined as coming from a content creator with whom the user has \emph{not} previously engaged. Training the ML model entails minimizing the following cross-entropy loss \citep{hastie2009elements} on the training data $\mathcal{D}$:
\begin{equation}
\label{eq:cross_entropy}
\begin{aligned}
\hat{\L} (\theta) = -\sum_{(x,y^v)\in\mathcal{D}} \sum_{v \in \mathcal{V}} y^v \log f (x; \theta).
\end{aligned}
\end{equation}

\begin{figure}[hbtp!]
    \centering
    \includegraphics[width=0.6\linewidth]{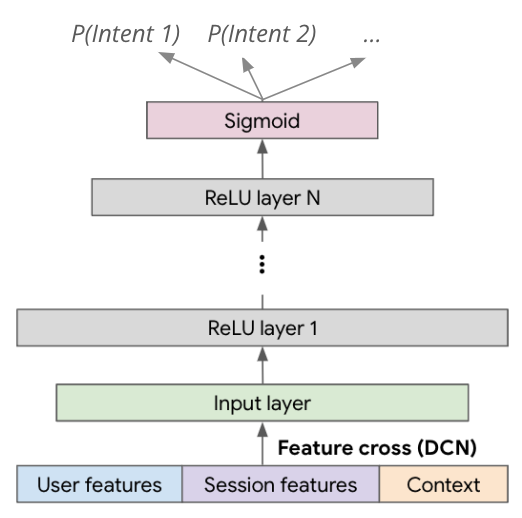}
    \caption{User Intent Model.}\label{fig:intent_model}
\end{figure}

In terms of the model architecture for $f^v(\cdot; \theta)$, we opt for a multi-layer neural network with a Deep \& Cross Network (DCN) \citep{wang2017deep} component for efficiently capturing the interactions among different features. Figure \ref{fig:intent_model} shows the architecture of the intent model.

\subsection{Intent Diversification Algorithm}
\label{sec:method_formulation}
With the components defined and derived in Section \ref{sec:method_components} and \ref{sec:user_intent_modeling}, we are now ready to derive the intent diversification algorithm. Our method is motivated by the search diversification algorithm proposed by \cite{agrawal2009diversifying} for diversifying search results for ambiguous queries. In the context of search, a query could belong to multiple categories and can have more than one interpretation \citep{radlinski2006improving}. The authors propose to greedily diversify the search results by selecting one result at a time that maximizes the probability of satisfying an average user, assuming the previously presented results do not satisfy them. We bring such logic to the recommendation space by mapping a query to a user, and mapping the query's category to the user's intent. Specifically, by law of total expectation, the probability that item $j$ satisfies user $i$, considering all the possible intents on a page is equal to: 

\begin{equation}
\label{eq:law_total_expectation}
\sum_{v \in \mathcal{V}} \Pr(v | i) Q (j | i, v),  
\end{equation}

In a search diversification framework such as \cite{agrawal2009diversifying}, the item that maximizes Eq.(\ref{eq:law_total_expectation}) is selected as the top search result. However, it is not straightforward to directly apply such search diversification algorithm to the recommendation context. Unlike the search context, where relevance typically serves as the single objective, recommender systems involve multiple objectives summarized into a single \emph{quality score}, $s_{ij}$, derived from various machine learning model outputs. As a result, this score often lacks a probabilistic interpretation. To integrate such non-probabilistic information into a probabilistic framework, we propose to multiply the quality score with the probabilistic intent-based prediction in Eq.(\ref{eq:law_total_expectation}). Specifically, at the first step of the intent diversification algorithm, we define the item at the top position, $j_1$ as 
\begin{equation}
\label{eq:j_1}
j_1 = \argmax_{1\leq j \leq M} \left\{ s_{ij} \cdot \left\{ \sum_v \Pr(v | i) Q (j | i, v)  \right) ^ \gamma \right\},
\end{equation}

where $\gamma > 0$ is the hyperparameter controlling the strength of intent diversification component. We opt for a multiplicative design over an additive one in combining the quality score with the intent-based component. This is because in contrast to an additive design, our proposed multiplicative approach offers larger boost to items having \emph{both} a higher intrinsic value $s_{ij}$ (without considering user intent) \emph{and} a higher intent value. This enables users to discover high-value items that also align with their intent, toward achieving an improved long-term experience on the platform.

Starting from the second position, the choice of which item to place at each subsequent position is influenced by the items already placed above that position. Imagine a user who examines the top 2 positions in order to make a decision whether to consume either item or quit. The objective of the recommender system is to maximize the value from \emph{either} of the top 2 items. Intuitively, this implies that deciding what to put at the second position involves a \emph{counterfactual} consideration, assuming that the user is \emph{not} interested in the first item. Therefore, the second position should pick an item that is \emph{less} likely to cover the intent that the first item has.

As an example, suppose the first position is a yoga item (i.e. $j_1$ is aligned with yoga intent). When deciding what to put at the second position, we would assume that the imaginary user is \emph{not} interested in the first yoga item. This implies that our belief about the imaginary user having a yoga intent should probably decrease, as she ``did not'' like the yoga item placed in the first position; however, the probability estimate of yoga intent should not drop to zero, as there is a chance that the user does have a yoga intent, just that the presented yoga item is not interesting to her. As a result, the second item should be \emph{less} likely to be a yoga-related item.

The example above suggests that a \emph{counterfactual} update of the user intents can be leveraged to solve for the intent-based diversification. Specifically, at step 2, we first compute the \emph{counterfactual posterior intent updates} based on the assumption that the user is \emph{not} interested in (rejects) the item at the top position ($R1 = (j_1)$), denoted as $\Pr(v | i, R_1)$:
\begin{equation}
\label{eq:counterfactual_intent}
\begin{aligned}
\Pr(v | i, R_1)  = 
\begin{cases}
 \frac{(1 - Q(j_1 | i, v) ) \Pr(v | i) }{ 1 - Q(j_1 | i) }, & \text{if } v \in \mathcal{V}_{j_1}, \\
 \Pr(v | i), & \text{otherwise},
\end{cases}
\end{aligned}
\end{equation}
where again $\mathcal{V}_{j_1}$ is the set of intents that item $j_1$ aligns to. Eq.(\ref{eq:counterfactual_intent}) is obtained by applying Bayes' theorem. Given the updated intent estimates, we now select the item for the second position as 
\begin{equation}
\label{eq:j_2}
j_2  = \argmax_{1\leq j \leq M} \left\{ s_{ij} \cdot \left(\sum_v \Pr(v | i, R_1) Q (j | i, v) \right) ^ \gamma \right\}.
\end{equation}
At this point, we have formulated a greedy algorithm, as determining the ranking for position 3 and beyond follows the same exact logic: at position $m$, we select an item $j_m$ to maximize the the following quantity: 
\begin{equation}
\label{eq:j_m}
j_m  = \argmax_{1\leq j \leq M} \left\{ s_{ij} \cdot \left(\sum_v \Pr(v | i, R_{m-1}) Q (j | i, v) \right) ^ \gamma \right\},
\end{equation}
by assuming that the user is \emph{not} interested in the preceding $m-1$ items (i.e., $R_{m-1} = (j_1, ... , j_{m-1})$).

Formally, the intent diversification algorithm described above is detailed in Algorithm \ref{algo:intent_diversification} below. On a high level, the algorithm initiates with $R_0 = \emptyset$, and iteratively selects one item $j_m$ at each step $m$ to be added to $R_{m-1}$ to formulate $R_{m}$ for $m = 1, \ldots, M$. It concludes after $M$ steps at $R_M = (j_1, \ldots, j_M)$, which serves as the final intent-diversified recommendation results. 

\begin{algorithm}
\caption{Intent Diversification for User $i$}\label{algo:intent_diversification}
\label{algo1}
\begin{algorithmic}[1]
   \REQUIRE $\Pr(v | i)$, $Q(j | i, v)$, $s_{ij}$ for $j \in \{1,\cdots,M\}$ and $v \in \mathcal{V}$, $\gamma$ 
   \ENSURE Intent-diversified recommendation list $(j_1,...,j_M)$ 
   \STATE $R_0 \leftarrow \emptyset$,  
   \STATE $S = \{1, \ldots, M\}$  
   \STATE $\Pr(v | i, R_0) \leftarrow \Pr(v | i)$   \COMMENT{prior intent belief}
   \FOR{$m = 1, \ldots, M$}  
       \FOR{$j \in S$}
           \STATE $Q(j | i, R_{m-1}) = \sum_{v \in \mathcal{V}} \Pr(v | i, R_{m-1}) Q(j | i, v)$ 
        \ENDFOR
        \STATE $j_m = \argmax_{j \in S} \left\{ s_{ij} \cdot Q(j | i, R_{m-1})^\gamma \right\}$      \COMMENT{item at position $m$}
        \STATE $R_m = R_{m-1} \oplus (j_m)$                  
        \STATE $S \leftarrow S \setminus \{j_m\} $   
        \STATE 
        $\Pr(v | i, R_m) = 
        \begin{cases}
          \frac{\Pr(v | i, R_{m-1}) (1 - Q(j_m | i, v) )}{ (1-Q(j_m | i))}, & \text{if } v \in \mathcal{V}_{j_m}, \\
         \Pr(v | i, R_{m-1}), & \text{otherwise},
        \end{cases} $
        \label{algo:posterior_intent} \\
        \COMMENT{posterior intent belief}
        
   \ENDFOR
   \RETURN $R_M = (j_1, \ldots, j_M)$   \COMMENT{final ranking}
\end{algorithmic}
\end{algorithm}

\paragraph{Understanding the algorithm.}

\begin{figure*}[hbtp!]
    \centering
    \includegraphics[width=1.0\linewidth]{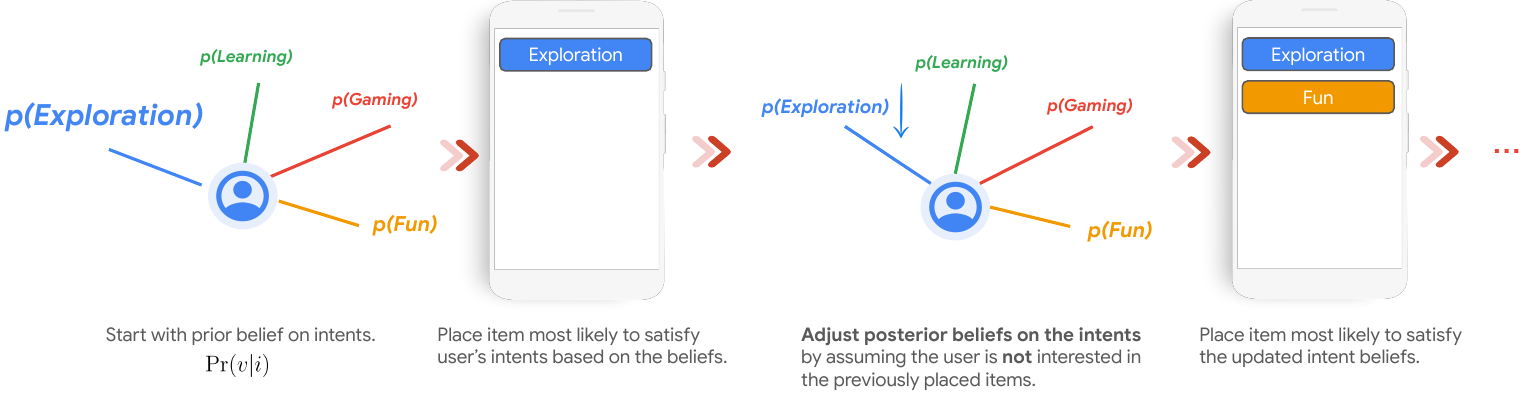}
    \caption{An illustration of the proposed intent diversification framework. The font size assigned to different intents visually represents their relative significance of their values. }\label{fig:overview}
\end{figure*}

Figure \ref{fig:overview} shows an intuitive illustration of the proposed intent diversification algorithm.  
As an example, user exploration intent starts with the highest propensity $p(Exploration)$ among the four intents (exploration, learning, gaming, fun). As a result, the top item selected is likely to be an exploratory item.\footnote{Note that it is not necessarily true that the top selected item is guaranteed to be an item from the intent with highest propensity. Instead, it also depends on the quality of the item according to Eq.(\ref{eq:j_m}).} Subsequently, the posterior belief on exploration intent drops because in the second position, we assume that the user is \emph{not} interested in the first item, leading to a \emph{decreased} posterior belief on the exploration intent. As a result, the relative importance of the rest of the intents (learning, gaming, fun) will \emph{increase}, therefore more likely to show up in the subsequent items. Intuitively, by assuming that the user is not interested in previously placed items, this greedy and sequential diversification procedure ensures that all intents are well represented in the recommendation. 
The multiplication of the quality score $s_{ij}$ further ensures that the item selected at each position maintains high relevance to the users.

\section{Experiments}
\label{sec:experiments}

\subsection{Experiment Setup}

We conduct experiments on YouTube, the world's largest video recommendation platform, serving billions of users everyday. Similar to many industrial recommendation platforms, the recommender system adopted by the platform for the landing page consists of three stages, including candidate generation, multi-tasking ranking, and whole-page optimization \citep{yi2023online, ma2018modeling, wilhelm2018practical}. Our intent diversification framework is applied as the last step in the whole-page optimization stage, taking the scores from the original multi-stage system as $s_{ij}$. Using live A/B testing, we compare the performance of the production recommender system with (treatment) and without (control) our proposed intent diversification framework. The experiments were run for a period considered long term by the company\footnote{Unfortunately, we cannot disclose the exact time period due to business compliance reasons.}, allowing us to observe the effects on long-term user experience.

While our proposed intent diversification framework works with any number or types of intents, in the first set of live experiments we selected two intents that are critical to the platform: \emph{exploration intent} and \emph{familiarity intent}. Following the intent label definition in Section \ref{sec:user_intent_modeling}, a user is considered to have an exploration intent if they prefer to consume content from an unseen creator (i.e., watch a video from a channel they have not seen before), and a familiarity intent if they prefer content from familiar creators (i.e., watch a video from channels they have seen before). The reason we selected these two intents is that they are closely connected with user exploration in recommender systems, which is defined as identifying unknown user interests or introducing users to new interests \citep{herlocker2004evaluating}. \citet{chen2021values} showed that user exploration is tightly connected to long-term user experience on content recommendation platforms. 
In addition, incorporating a user's exploration intent in the diversification stage allows the system to dynamically adjust the diversity level based on whether the user seeks explorational content. This approach enables the diversification stage to adapt to user preferences in real time, effectively implementing a form of ``learning to diversify'' in recommender systems. Motivated by these considerations, we chose user exploration and familiarity intents for our intent diversification framework to optimize long-term user experience. The intent model is retrained every 6 hours with the latest observations to capture any dynamics in the intent space. 

In Section \ref{sec:exploration_intent_results} below, we present live A/B testing results on these intents. In Appendix \ref{sec:appen_add_intent}, we present live A/B testing results on a different set of intents, specifically creator-level intents and visit length intents, to showcase the generalizability and scalability of our proposed framework to a large number of intents.

\subsection{Live Experiment Results}
\label{sec:exploration_intent_results}
\begin{figure}[hbtp!]
    \begin{subfigure}[b]{0.22\textwidth}
        \centering
        \includegraphics[width=\textwidth]{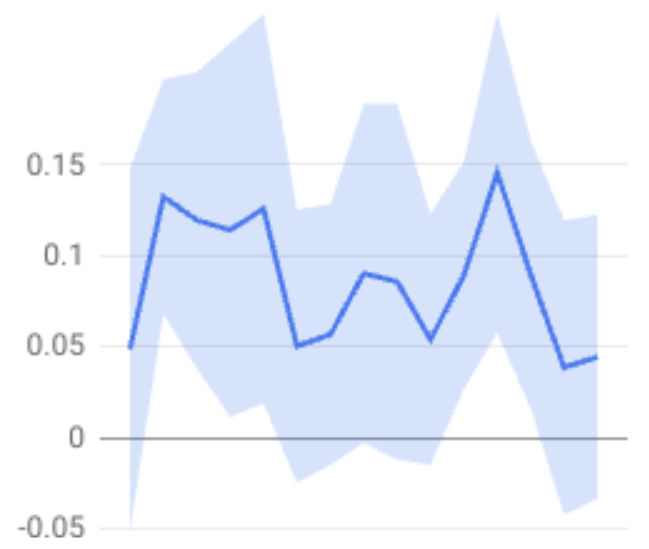}
        \caption{Overall user enjoyment.}
        \label{fig:overall_enjoyment}
    \end{subfigure}
    \hspace{3mm}
     \centering
    \begin{subfigure}[b]{0.22\textwidth}
        \centering
        \includegraphics[width=\textwidth]{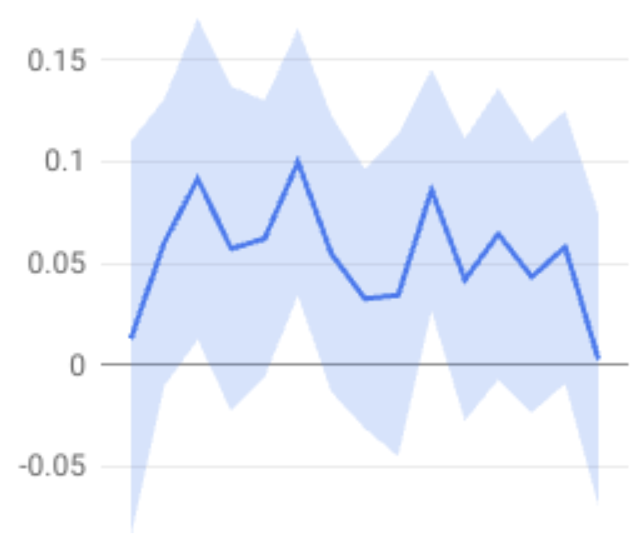}
        \caption{Daily Active Users.}
        \label{fig:dau}
    \end{subfigure}
    \hspace{3mm}
     \centering
    \begin{subfigure}[b]{0.22\textwidth}
        \centering
        \includegraphics[width=\textwidth]{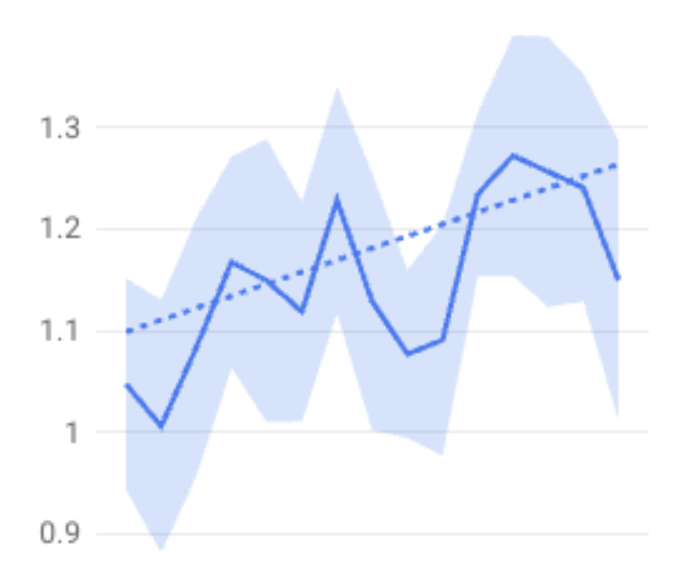}
        \caption{Landing page consumption.}
        \label{fig:home_consumption}
    \end{subfigure}
    \hspace{3mm}
    \begin{subfigure}[b]{0.22\textwidth}
        \centering
        \includegraphics[width=\textwidth]{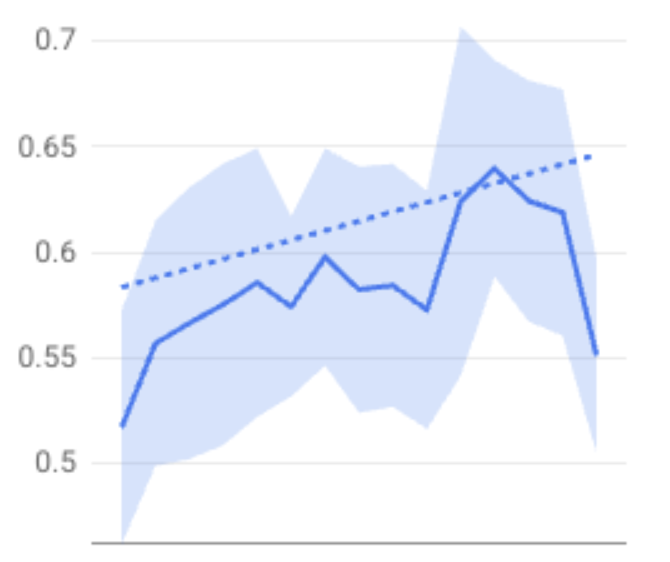}
        \caption{Satisfied users.}
        \label{fig:satisfied_home_watcher}
    \end{subfigure}
   \caption{Live experiment results on the intent diversification framework. Results are reported as percentage changes over the control group over the experiment period.}
  \label{fig:live_exp}
\end{figure}

Figure \ref{fig:live_exp} summarizes the live experiment metrics with the selected exploration and familiarity intents. The differences are reported as percentage changes on the metrics of the treatment group with respect to the control. Due to business compliance reasons, the absolute values of the metrics and the x-axis (representing the time horizon of the experiment) have been omitted. The top-line business metric, overall user enjoyment\footnote{This is an \emph{objective} and well-defined measure on high-quality user engagement on the platform. It combines feedback based on time spent, survey, and content quality guardrail etc. We have kept the name of the metric vague due to business compliance reasons.}, shows a significant improvement of 0.09\% (Fig.\ref{fig:overall_enjoyment}) under the intent diversification treatment. At the same time, Daily Active Users (DAU) is improved by 0.05\% which is also statistically significant (Fig.\ref{fig:dau}). This is accompanied by the upward trend in the improvement in  consumption on the landing page (Fig.\ref{fig:home_consumption}), indicating that the users are increasingly engaging with the platform. Furthermore, there is also an upward trend in satisfied users (Fig.\ref{fig:satisfied_home_watcher}), suggesting that users are not only increasing their engagement with the platform but also expressing higher satisfaction levels. The improvement in DAU suggests that users are returning to the platform more frequently, implying an improved long-term user experience.

We emphasize that while the absolute percentages may appear small, they represent a significant business impact considering the scale of the platform. Notably, on large platforms, DAU movement has reached saturation, making it increasingly difficult to observe any statistically significant changes in DAU. 

\subsection{Understanding the Improvements}

\subsubsection{Overall exploration and diversity patterns.} 
\begin{figure}[hbtp!]
    \begin{subfigure}[b]{0.22\textwidth}
        \centering
        \includegraphics[width=\textwidth]{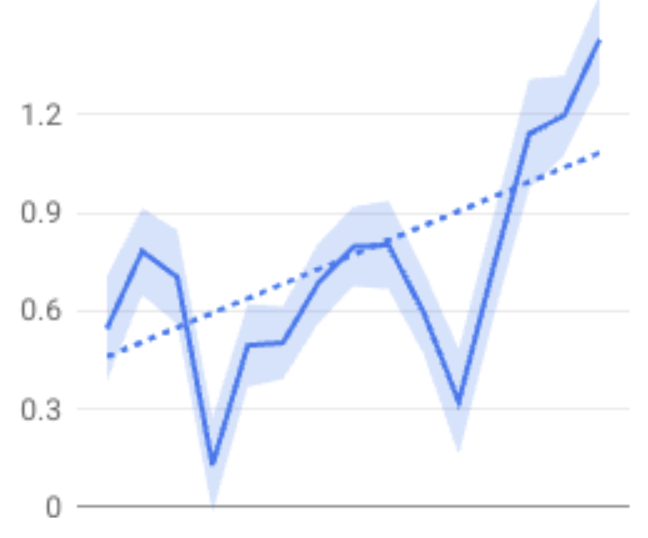}
        \caption{Novel content creator engagement.}
        \label{fig:novel_channel_clicks}
    \end{subfigure}
    \hspace{3mm}
     \centering
    \begin{subfigure}[b]{0.22\textwidth}
        \centering
        \includegraphics[width=\textwidth]{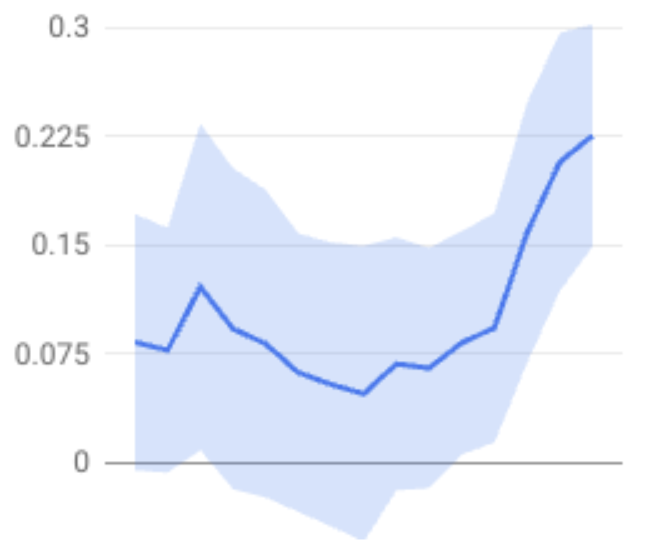}
        \caption{Repeated Explorations.}
        \label{fig:repeated_explore}
    \end{subfigure}
    \hspace{3mm}
     \centering
    \begin{subfigure}[b]{0.22\textwidth}
        \centering
        \includegraphics[width=\textwidth]{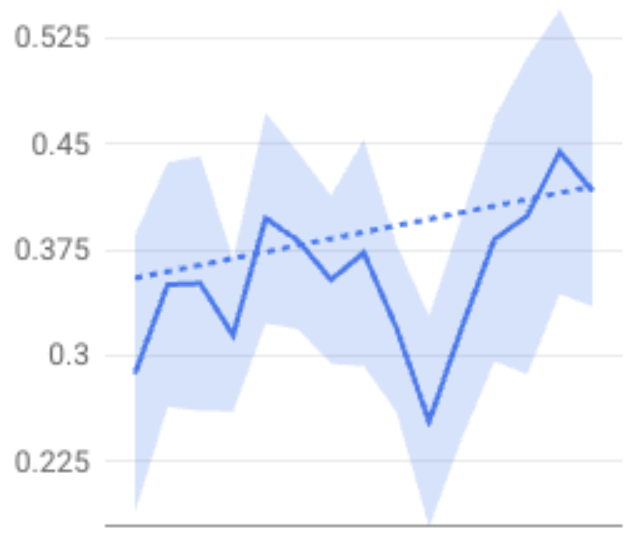}
        \caption{Average unique clusters consumed per user.}
        \label{fig:unique_cluster}
    \end{subfigure}
   \caption{Exploration and diversity related metrics from the live experiment. Results are reported as percentage changes over the control group over the experiment period.}
  \label{fig:exploration_diversity_metric}
\end{figure}

To delve deeper into the sources of gain from the intent diversification framework, we conducted additional analyses to examine how users' exploration behavior evolves over time and its impact on the diversity of their consumption patterns. In addition to the improvements in top business metrics in Section \ref{sec:exploration_intent_results}, we also see a significant improvement in user engagement with novel creators (Fig.\ref{fig:novel_channel_clicks}). To check whether these engagements are high-quality engagements, the platform also measures ``repeated exploration'', which is defined as the user visiting the same novel creator multiple times. This can be viewed as a measure for high-quality exploration behavior. Figure \ref{fig:repeated_explore} shows that there is also a significant improvement in repeated exploration, indicating an improvement in users finding high-quality exploratory contents. To measure the diversity of users' consumption patterns, we also examined the unique topic clusters that users interacted with\footnote{These topic clusters, derived from an unsupervised clustering algorithm based on observed co-occurrence patterns on the platform \citep{chen2021values}, indicate semantic similarities between items within the same cluster.}. Notably, Fig.\ref{fig:unique_cluster} demonstrates a substantial improvement in the average number of unique topic clusters consumed per user. This finding suggests that the intent diversification strategy, by encouraging users to engage with a wider variety of novel content, leads to increased consumption diversity.

Importantly, Fig.\ref{fig:exploration_diversity_metric} demonstrates a notable upward trend across all three metrics, suggesting increasing improvements in exploration and diversity over time. This trend implies that users are increasingly engaging with a wider set of exploratory and diversified content over their time spent on the platform. Prior research has highlighted the significant correlation between high-quality exploration and better long-term user experience \citep{chen2021values}. Building on this, the observed improvements in long-term user experience resulting from our intent diversification framework can be attributed to its facilitation of consistent content discovery and engagement content \emph{over time}. This continuous and increasing exposure to novel and diverse content contributes to a more satisfying and sustained user experience in the long run.

\subsubsection{The value of personalization in intent diversification.} 
\begin{figure}[hbtp!]
    \begin{subfigure}[b]{0.23\textwidth}
        \centering
        \includegraphics[width=\textwidth]{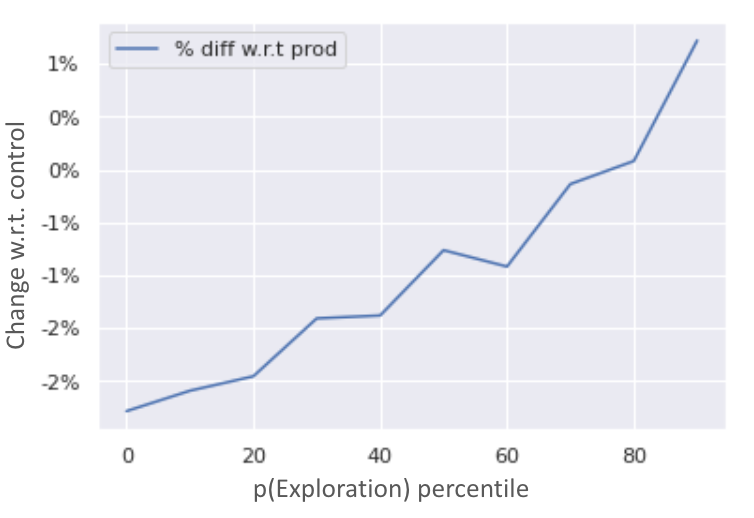}
        \caption{Novel impressions.}
        \label{fig:novel_impression_ratio}
    \end{subfigure}
     \centering
    \begin{subfigure}[b]{0.23\textwidth}
        \centering
        \includegraphics[width=\textwidth]{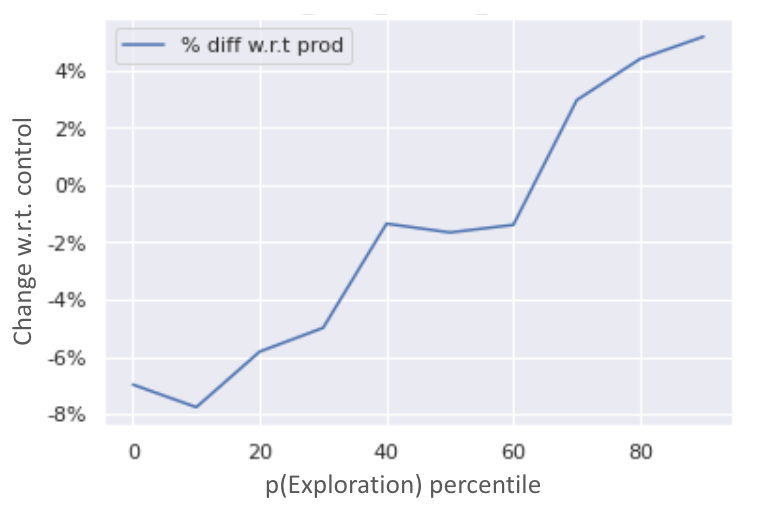}
        \caption{Novel consumptions.}
        \label{fig:novel_consumption_ratio}
    \end{subfigure}
     \centering
    \begin{subfigure}[b]{0.23\textwidth}
        \centering
        \includegraphics[width=\textwidth]{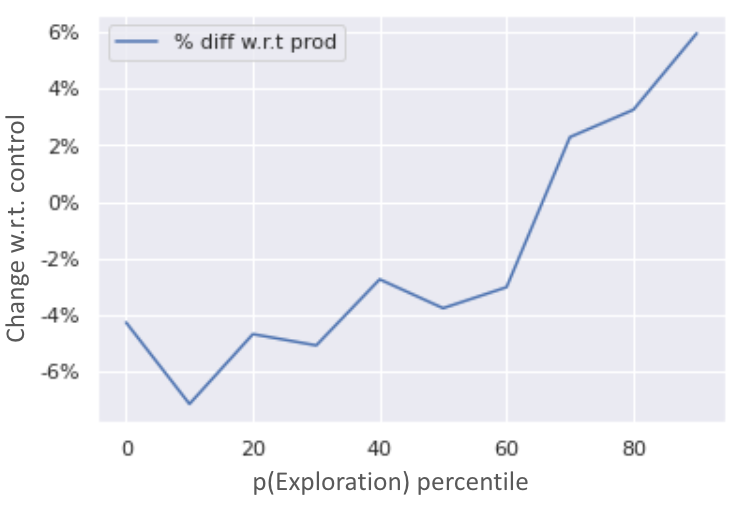}
        \caption{Novel CTR.}
        \label{fig:novel_ctr}
    \end{subfigure}
   \caption{Comparison between treatment and control, sliced by percentile of the predicted exploration intent. Results are reported as percentage changes over the control group.}
  \label{fig:sliced_by_percentile}
\end{figure}

Our proposed intent diversification framework is a personalized treatment based on a personalized prediction of intent propensities for each user. To understand the effect of personalization, we conducted further analysis by comparing the treatment (with intent diversification) and control (without intent diversification) for users with different levels of exploration intent $p(Exploration)$.\footnote{As the two intent probabilities sum to one, our analysis focuses only on $p(Exploration)$ here; The findings for the familiarity intent are symmetric.} In particular, we looked at the following metrics: (1) \emph{Novel impressions}: Number of impressions on the landing page that are from novel content creators (channels), with whom the user has not previously engaged.\footnote{Note that the novelty of content creators is a personalized notion. A novel content creator for user A may not be a novel content creator for user B.} This measures the overall novelty of the recommendations. (2) \emph{Novel consumptions}: Number of consumed items on the landing page that are from novel content creators. (3) \emph{Novel click-through rate (CTR)}: Click-through-rates of the novel recommended items. This measures the relevance of the novel recommendations. Figure \ref{fig:novel_impression_ratio} - \ref{fig:novel_ctr} shows the differences between treatment and control in these metrics. We observe that for pages with higher exploration intent (and consequently lower familiarity intent), the treatment (our intent diversification framework) is recommending \emph{more} novel items (Fig.\ref{fig:novel_impression_ratio}) than control, and the users are more likely to engaging with those novel contents (Fig.\ref{fig:novel_consumption_ratio} and Fig.\ref{fig:novel_ctr}). 

All three metrics in Fig.\ref{fig:sliced_by_percentile} exhibit a consistent monotonic trend. This suggests that our intent diversification framework improves personalization with respect to users' intents. Specifically, for pages with a higher propensity towards exploration, our framework not only recommends an increasing proportion of high-quality novel items but also observes increased user engagement with these recommendations, as evidenced by the difference in CTR (Fig.\ref{fig:novel_ctr}). For example, at about 80\% percentile of $p(Exploration)$, our proposed framework is showing about the same amount of novel recommendations (Fig.\ref{fig:novel_impression_ratio}), but results in 4\% increase in consumptions from novel contents (Fig.\ref{fig:novel_consumption_ratio}). This suggest that our framework is more efficient in recommending contents that are both novel and relevant to the users who are in exploration mode. Conversely, for users with lower exploration intent, the recommendation of such novel items decreases and the users are less likely to engage with them. In other words, our intent diversification framework is able to further personalize the recommendation results to align them with varying levels of user exploration intents. In Appendix \ref{sec:appen_exploration}, we also discuss comparisons between our framework and additional baselines on exploration.  


\subsection{Understanding User Intents}

\subsubsection{Model performance and feature correlation analysis}
\begin{figure}[hbtp!]
    \centering
    \includegraphics[width=0.7\linewidth]{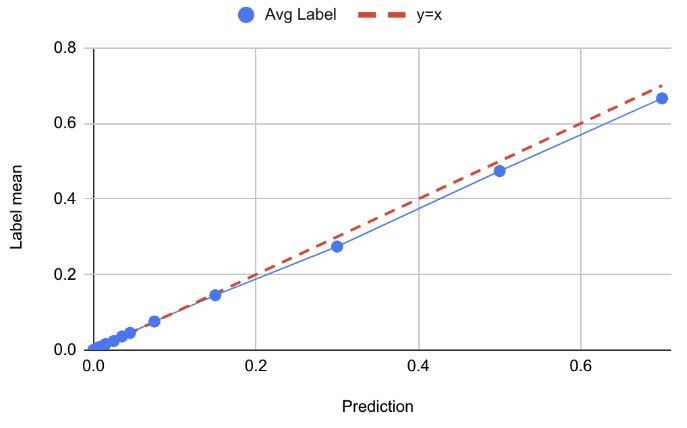}
    \caption{Calibration plot of the user exploration intent model. A perfectly calibrated model has a calibration score of 1 and the calibration curve aligns with the 45 degree line.}\label{fig:calibration}
\end{figure}

As the probabilities of exploration intent and familiarity intent sum to one, we only present the analysis for exploration intent here; the results for familiarity intent are symmetric. The user intent model has an AUC of 0.73 and is well calibrated with an average prediction to label ratio of 0.97 (Fig.\ref{fig:calibration}). To understand what signals are captured by the user intent model, we conducted analysis to study how each feature is correlated with the $p(Exploration)$ prediction and rank by the order of the absolute value of the correlations. Table \ref{tab:feature_cor} summarizes the features that have the highest absolute correlations with the predicted user exploration intent. We see that features such as ``current session length'' and ``number of consumptions in the current session'' are \emph{positively} correlated with $p(Exploration)$. This is interesting and expected, as it suggests that users are more likely to explore when they have caught up existing interests (e.g. finishing consuming the newest episode of a drama series) in the current session. Features such as ``average completion ratio of past consumed items'' and ``average total length of past consumed items'' are negatively correlated with $p(Exploration)$, indicating that users who are less patient with each consumption (i.e. more likely to consume shorter items or quit before finishing the consumption) are more likely to explore. These findings are also aligned with the qualitative insights from user interviews conducted at the company.

\begin{table}[htbp!]
  \begin{center}\footnotesize
   \setlength\extrarowheight{4pt}
    \begin{tabular}{|l|}
    \hline
       \textbf{Features that are positively correlated with $p(Exploration)$}    \\ \hline
       Number of topic clusters consumed \\
       Number of content creators interacted with \\ 
       Current session length \\ 
       Number of consumptions in the current session \\ 
       Time since last session \\
       Past consumption diversity \\
       \hline
       \textbf{Features that are negatively correlated with $p(Exploration)$}  \\ \hline
       Average completion ratio of past consumed items\\ 
       Average total length of past consumed items \\
       Average consumption time \\ 
       Past activity level \\
       Number and ratio of repeated consumptions  \\
       \hline
    \end{tabular}
    \caption{Feature analysis of user intent model.}
    \label{tab:feature_cor} 
  \end{center}
\end{table}


\subsubsection{User intents on different time horizons}

\begin{figure}[H]
    \begin{subfigure}[b]{0.22\textwidth}
        \centering
        \includegraphics[width=\textwidth]{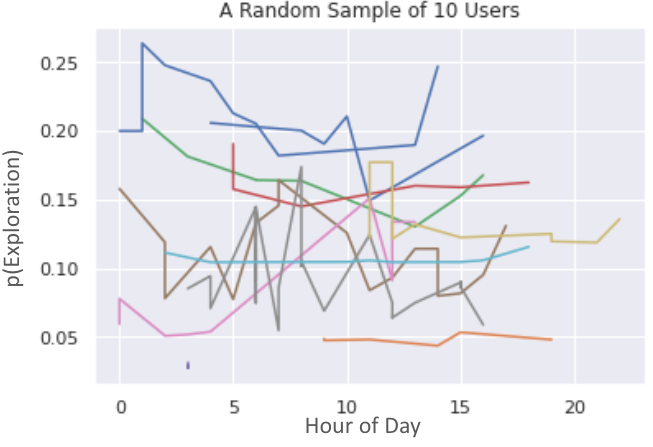}
        \caption{Page-level exploration intent: $p(Exploration)$ over different hours of day.}
        \label{fig:pexplore_hour}
    \end{subfigure}
    \hspace{0.5mm}
     \centering
    \begin{subfigure}[b]{0.22\textwidth}
        \centering
        \includegraphics[width=\textwidth]{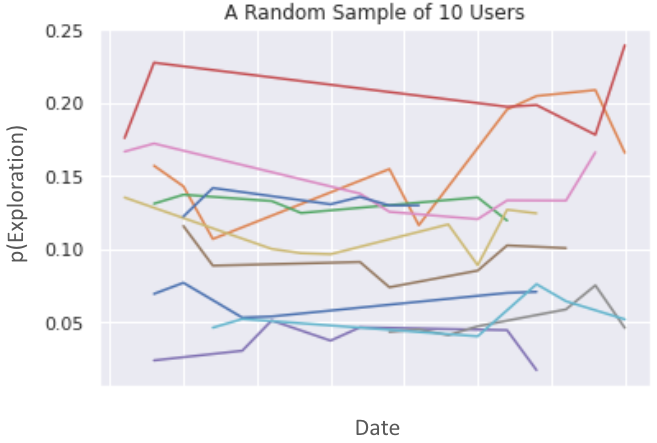}
        \caption{User-level exploration intent: Average $p(Exploration)$ over different days.}
        \label{fig:pexplore_day}
    \end{subfigure}
   \caption{User intents on different time horizons.}
  \label{fig:pexplore_hierarchical}
\end{figure}

To understand how the predicted user intents evolves over different sessions and varies across different users, we visualized one of the intent model's predictions $p(Exploration)$ across different time horizons for a random sample of the users. Figure \ref{fig:pexplore_hour} shows the page-level $p(Exploration)$ over different hours of day for a random sample of 10 users. We observe that users tend to exhibit varying levels of exploration intent throughout different hours or sessions of the day, even for the same individual. This is expected, as a user's mood and context may fluctuate between, for instance, the morning rush hours and evening family time, resulting in differing propensities to explore. Figure \ref{fig:pexplore_hour} shows that our intent model is able captures these variations. 

Even more intriguingly, when aggregating data at the daily level rather than hourly, we observe significantly smaller fluctuations across different days, as shown in Fig.\ref{fig:pexplore_day}. This implies that each user maintains a relatively stable baseline level of exploration intent, which may vary among individuals. While some users are inherently more likely to explore than others, within the same user, their propensity to explore remains consistent across different days. This stable nature of user-level intent provides grounds in using supervised machine learning models to predict future intents based on the user's past behaviors.

\subsection{Effect of the Strength of Intent Diversification}
Although diversity is a desired property for recommendation results, it has been established that diversity usually hurts the overall relevance of the set of recommended items \citep{chen2021values}. Therefore, the goal of the recommender system is to achieve the optimal trade-off between diversity and relevance in order to optimize long-term user experience. In our intent diversification framework, the strength of diversification is controlled by the hyperparameter $\gamma$ in Eq.(\ref{eq:j_m}). To understand the effect of the different strengths of intent diversification, we conduct additional experiments with varying values of $\gamma$, and measure the following four metrics: (1) \emph{Diversity}, measured by number of unique clusters consumed per user; (2) \emph{Novelty}, measured by novel content creator engagement; (3) \emph{Relevance}, which is the average of the relevance score (predicted by a separate ML model) of all recommended items on the landing page; (4) \emph{Long-Term Metric (DAU)}, which is the Daily Active User (DAU) metric. 

\begin{figure}[htbp!]
    \centering
    \includegraphics[width=0.8\linewidth]{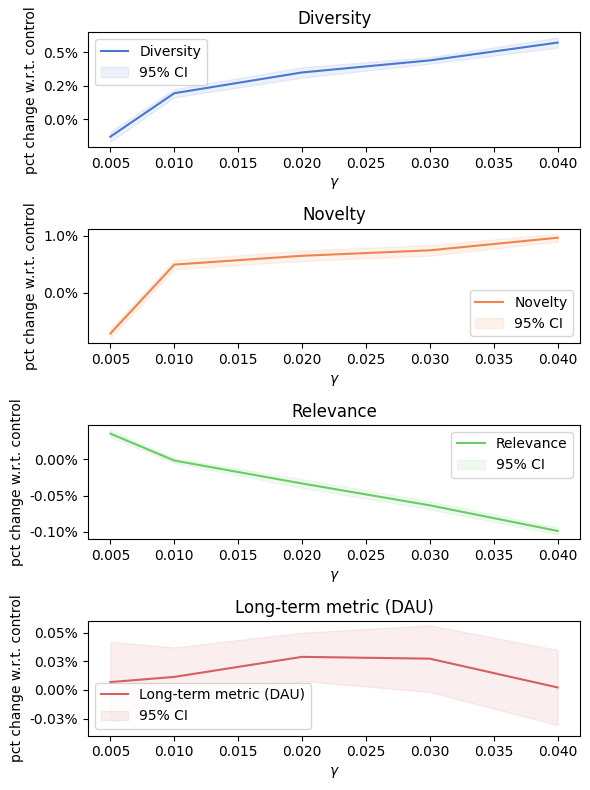}
    \caption{Changes in diversity, novelty, relevance and long-term metric for different strengths ($\gamma$) of intent diversification. Metrics are reported as relative changes compared to control which does \emph{not} have intent diversification.}\label{fig:tune_gamma}
\end{figure}

Figure \ref{fig:tune_gamma} shows how these four metrics change with different strength levels of intent diversification. We see that as the strength $\gamma$ increases from 0.005 to 0.04, there is a monotone increase in diversity and novelty, and a monotone decrease in relevance (top three subplots of Fig.\ref{fig:tune_gamma}). This confirms the inherent trade-off between diversity and relevance and shows that our intent diversification framework serves as an effective lever to control this trade-off. The last plot in Fig.\ref{fig:tune_gamma} shows a non-monotone trend of DAU, which is maximized around $\gamma = 0.02$. The live experiment results in the previous sections are reported at $\gamma = 0.02$. At this point, neither diversity nor relevance reaches their maximum values. This shows that a delicate \emph{balance} between diversity and relevance of the recommender systems is necessary in order to optimize long-term outcomes (e.g. DAU) on these platforms.


\section{Conclusion}
\label{sec:conclusion}

In this work, we propose a framework to diversify recommendation results based on user intents. By adapting an algorithm originally designed for search results diversification, we propose to explicitly incorporate the user's propensities to different intents in the final stage of a recommender system, to ensure that different intents are represented in the recommendations. Through live experiments on YouTube, the world's largest video recommendation platform, we demonstrate the effectiveness of our proposed framework in assisting users towards their intents and improving their long-term experience on the platform. In particular, we see a significant improvement in long-term metrics such as overall user enjoyment and Daily Active Users (DAU). 

Our proposed framework works with any number of intents and can be applied either as a standalone diversification layer or combined with existing whole-page optimization algorithms. Although our findings is only from one platform, our framework is general and readily applicable to other recommendation platforms. In fact, our approach is more beneficial when there are \emph{not} a huge amount of training data available. This is because compared to a black-box diversification algorithm, our intent-based diversification framework is able to better capture the underlying data generation process (i.e. an intent-driven decision-making process). It improves learning efficiency given the same amount of training data as a result. Therefore, smaller platforms should benefit even more from our framework. In addition, the intent prediction model is much smaller in size compared with the other models used by the platform. Therefore, it introduces minimal engineering cost in implementing our framework.

In the live experiments, we selected the intents based on business considerations by the platform. A future research direction is to automatically learn a taxonomy of user intents that can directly serve as input to the intent diversification framework toward optimizing long-term user experience on the platform.

 %


\bibliographystyle{ACM-Reference-Format}
\bibliography{reference}

\appendix

\section{Additional experiment results}

\subsection{Experiments on additional intents}
\label{sec:appen_add_intent}

We define a creator intent as the user's propensity to engage with content from a specific creator on the platform (i.e., a channel on YouTube). Considering the fact that YouTube hosts hundreds of millions of creators, the set of intents $\mathcal{V}$ is extremely large in this case. We were able to conduct A/B testing on the intent diversification framework with the creator intents. 

Figure \ref{fig:channel_live_exp_intent_metrics} below summarizes the results. There is a significant 0.05\% increase in Daily Active Users (DAU) from incorporating creator intents (Fig.\ref{fig:channel_dau}). This is accompanied by a 0.04\% increase in site-wide consumption count (Fig.\ref{fig:channel_sitewide_click}). In terms of intent-related metrics, we see that there is a significant 0.56\% increase in novel creator consumption, 0.10\% increase in unique user-creator pairs, 0.13\% increase in unique user-cluster pairs, and 0.30\% repeated novel consumption, suggesting that our creator-intent-aware treatment facilitates the discovery and engagement with novel content creators. As a result, we also see an increase in consumption diversity which is measured by unique user-cluster pairs. 

\begin{figure}[hbtp!]
     \begin{subfigure}[b]{0.22\textwidth}
        \centering
        \includegraphics[width=\textwidth]{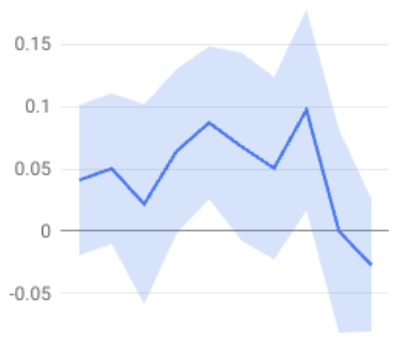}
        \caption{Daily Active Users.}
        \label{fig:channel_dau}
    \end{subfigure}
    \centering
    \begin{subfigure}[b]{0.22\textwidth}
        \centering
        \includegraphics[width=\textwidth]{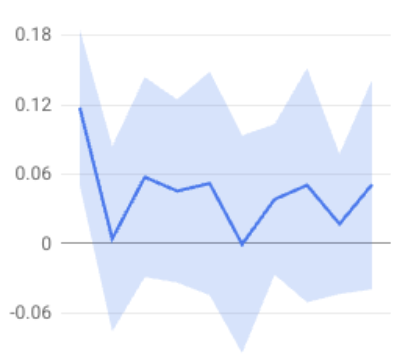}
        \caption{Site-wide consumption count.}
        \label{fig:channel_sitewide_click}
    \end{subfigure}
    \begin{subfigure}[b]{0.22\textwidth}
        \centering
        \includegraphics[width=\textwidth]{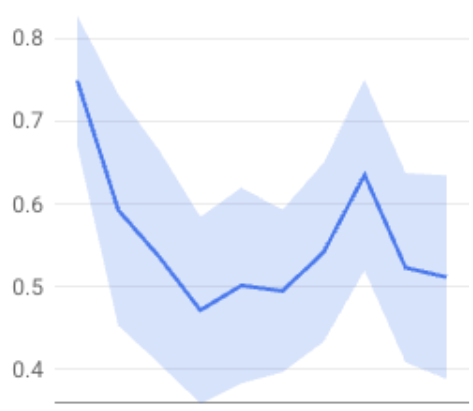}
        \caption{Novel creator consumption.}
        \label{fig:channel_novel_channel_clicks}
    \end{subfigure}
    \begin{subfigure}[b]{0.22\textwidth}
        \centering
        \includegraphics[width=\textwidth]{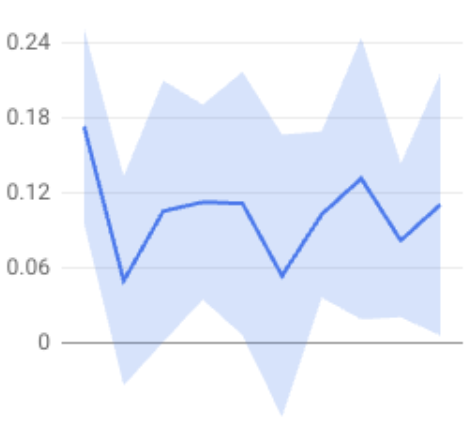}
        \caption{Unique user-creator pairs.}
        \label{fig:channel_unique_user_channel}
    \end{subfigure}
    \centering
    \begin{subfigure}[b]{0.22\textwidth}
        \centering
        \includegraphics[width=\textwidth]{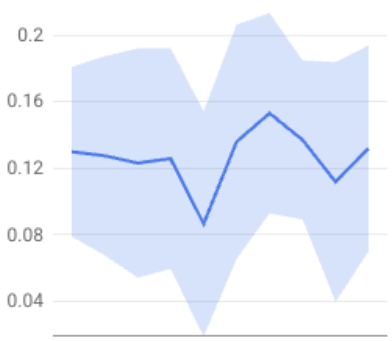}
        \caption{Unique user-cluster pairs.}
        \label{fig:channel_unique_user_cluster}
    \end{subfigure}
    \centering
    \begin{subfigure}[b]{0.22\textwidth}
        \centering
        \includegraphics[width=\textwidth]{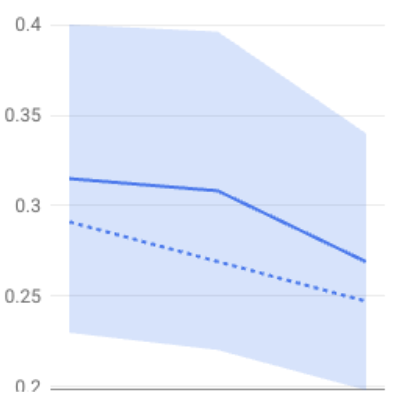}
        \caption{Repeated novel consumption.}
        \label{fig:channel_repeated_explore}
    \end{subfigure}
   \caption{Creator-intent related metrics.}
  \label{fig:channel_live_exp_intent_metrics}
\end{figure}

\begin{figure}[hbtp!]
    \begin{subfigure}[b]{0.22\textwidth}
        \centering
        \includegraphics[width=\textwidth]{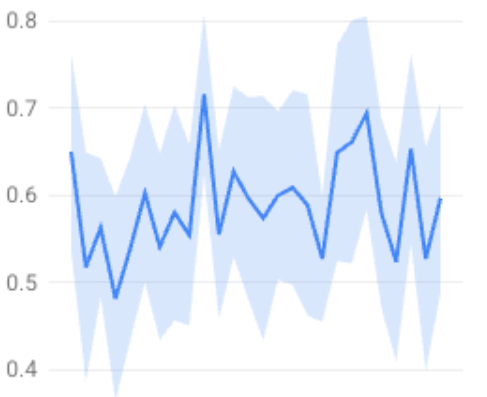}
        \caption{Overall user enjoyment.}
        \label{fig:visit_length_overall_enjoyment}
    \end{subfigure}
    \begin{subfigure}[b]{0.22\textwidth}
        \centering
        \includegraphics[width=\textwidth]{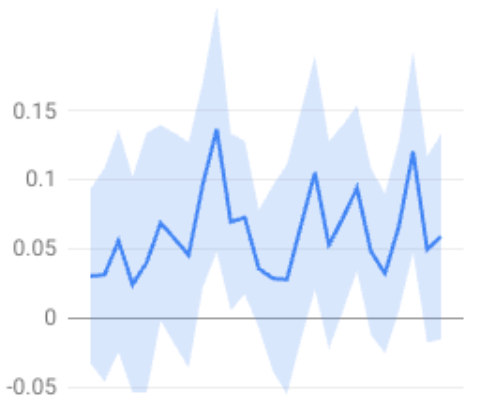}
        \caption{Landing page consumption.}
        \label{fig:visit_length_home_consumption}
    \end{subfigure}
   \caption{Visit length intent related metrics.}
  \label{fig:visit_length}
\end{figure}

We also observed positive A/B experiment results when modeling user intents around the visit length on the platform within our intent diversification framework. Specifically, we build an ML model to predict the probability that a user would prefer to watch a long versus a short video on the current recommendation page. By incorporating these visit length intents, we achieved a significant increase of +0.06\% in overall user enjoyment and a notable +0.59\% increase in landing page consumption, as shown in Fig.\ref{fig:visit_length}. These outcomes confirm our framework’s ability to effectively generalize and scale across a broad set of user intents.

\subsection{Additional baselines on exploration}
\label{sec:appen_exploration}

Our intent diversification framework with exploration and familiarity intents also outperforms existing state-of-the-art baselines for exploration in recommender systems. 
The existing production recommender system at the platform (which serves as control in our experiments) has already incorporated several state-of-the-art approaches for boosting novelty and diversity in the recommendation results. For diversification, the control group has incorporated a treatment similar to \citet{wilhelm2018practical} and \citet{wang2022surrogate}; For novelty, the control group has incorporated a treatment similar to \citet{chen2021values}. For general diversification algorithms, the platform has experimented with approaches ranging from rule-based heuristics to model-based diversification \citep{wilhelm2018practical}. Our proposed intent diversification algorithm is able to provide \emph{further} improvements (in DAU etc.) on top of these existing state-of-the-art approaches for diversified and novel recommendations.

\end{document}